\documentclass[prd,12pt,nofootinbib,preprint]{revtex4}
 
\usepackage{graphicx}
\usepackage{epsfig}  
\usepackage{epsf}    
\usepackage[tbtags]{amsmath}
\usepackage{amsfonts}
\usepackage{float}
\usepackage{subfig}
\usepackage{dcolumn}
\usepackage{bm}
\usepackage{dcolumn}
\usepackage{textcomp}
\usepackage{multirow}
\usepackage{slashed}
\usepackage{float}
\restylefloat{figure}
\usepackage[]{hyperref}
  \hypersetup{
  bookmarks=true,         
  unicode=true,         
  pdftoolbar=true,     
  pdfmenubar=true,     
  pdffitwindow=true,    
  pdfstartview={FitH},    
  pdfsubject={Sterilenus@DUNE},  
  pdfnewwindow=true,     
  pdfcreator={RevTeX},
  colorlinks=true,     
  linkcolor=red,   
  citecolor=blue,    
  filecolor=black,  
  urlcolor=blue,           
  }
\usepackage{hypcap}

\def\nue{{\nu_e}}
\def\anue{{\bar{\nu}_e}}
\def\numu{{\nu_{\mu}}}
\def\anumu{{\bar{\nu}_{\mu}}}

\newcommand{\beq}{\begin{equation}}
\newcommand{\eeq}{\end{equation}}
\newcommand{\beqa}{\begin{eqnarray}}
\newcommand{\eeqa}{\end{eqnarray}}

\begin{document}
\title{Capability of the proposed long-baseline experiments to probe large extra dimension}

\author{Samiran Roy}
\email[]{samroy@na.infn.it}
\affiliation{INFN - Sezione di Napoli, Complesso Univ. Monte S. Angelo, I-80126 Napoli, Italy}

\begin{abstract}
Future long-baseline experiments will play an important role in exploring physics beyond the standard model. One such new physics concept is the large extra dimension (LED), which provides an elegant solution to the hierarchy problem. This model also explains the small neutrino mass in a natural way. The presence of LED modifies the standard neutrino oscillation probabilities. Hence, the long-baseline experiments are sensitive to the LED parameters. We explore the potential of the three future long-baseline neutrino experiments, namely T2HK, ESSnuSB, and DUNE, to probe the LED parameter space. We also compare the capability of the charged and neutral current measurements at DUNE to constrain the LED model. We find that T2HK will provide more stringent bounds on the largest compactification radius ($R_{\rm{ED}}$) compared to the DUNE and ESSnuSB experiments. At $90\%$ C.L., T2HK can exclude $R_{\rm{ED}}\sim 0.45~(0.425)$ $\mu$m for the normal (inverted) mass hierarchy scenario.

\end{abstract}


\maketitle


\section{Introduction}
\label{sec1}
Neutrino oscillation firmly establishes the massive nature of neutrinos. The flavor and mass basis are not the same. They are related by a unitary mixing matrix which depends on three mixing angles ($\theta_{12},\theta_{13},\theta_{23} $) and one CP violating phase ($\delta_{cp}$). The bound on the absolute mass of the neutrino coming from cosmology~\cite{Planck:2018vyg} and direct detection experiment~\cite{KATRIN:2021uub} is an order of sub-electronvolt. The smallness of neutrino mass compared to other standard model (SM) particles remains a puzzle to the scientific community. There are many promising theories beyond standard model (BSM) that generate the small neutrino mass naturally. The inclusion of heavy right handed neutrinos produces the small neutrino masses via a see-saw mechanism \cite{PhysRevLett.44.912,PhysRevD.22.2227}. Another interesting mechanism that explains the small neutrino mass is the large extra dimension (LED) \cite{Arkani-Hamed:1998jmv,Antoniadis:1998ig,Arkani-Hamed:1998sfv} which was first introduced to solve the hierarchy problems, \textit{i.e.}, the large discrepancy between the electroweak scale ($M_{\rm{EW}}\sim 10^3$ GeV) and the Plank scale ($M_{\rm{Pl}} \sim 10^{19}$ GeV) where the interaction of gravity becomes strong.  In this model, the main assumption is that there exists only one fundamental scale which is the electroweak scale and in higher dimensions ($4+n$) the two scales become equivalent, \textit{i.e.}, $M_{\rm{Pl}} \sim M_{\rm{EW}}$. But in 4-dimensional space the value of $M_{\rm{Pl}}$ is large compared to $M_{\rm{EW}}$. The inclusion of $n$ compactified dimensions also modifies the inverse square law at a distance $10^{30/n~ -17}~({\rm{TeV}}/M_{\rm{EW}})^{(1+2/n)}$ cm~\cite{Arkani-Hamed:1998jmv}. One extra compactified dimension ($n=1$) case is ruled out from the observation of verified gravitational inverse square law at solar scale, while $n\geq 2$ scenario is allowed. Here, we consider an asymmetric space. Only one extra spatial dimension is large out of $n$ compactified dimensions, and the space is effectively 5-dimensional. In this model, all the SM particles are confined to the four dimensional brane which is a subspace of the full space-time. However, the SM singlet right handed neutrino could propagate in more than 4-dimensions and the large volume of extra dimensions provides suppression to the field in the 4-dimensional space. This makes the mass of a neutrino very small~\cite{Arkani-Hamed:1998wuz,Dienes:1998sb,Dvali:1999cn,Barbieri:2000mg,Nortier:2020lbs} compared to other SM particles in a natural way.

The next generation long-baseline experiments will play an important role in measuring the oscillation parameters with percentage-level precision. We consider three future long-baseline experiments, namely T2HK~\cite{Hyper-KamiokandeProto-:2015xww,Hyper-Kamiokande:2016srs,Hyper-Kamiokande:2018ofw}, DUNE~\cite{DUNE:2015lol,DUNE:2016rla}, and ESSnuSB~\cite{ESSnuSB:2013dql,Alekou:2022emd,Ghosh:2021fua}. The primary goal of these experiments is to address issues like mass hierarchy ($\Delta m^2_{31}>0$ normal ordering, $\Delta m^2_{31}<0$ inverted ordering), the octant of atmospheric mixing angle ($\theta_{23}$), the determination of the value of $\delta_{cp}$ etc. These experiments are also sensitive to physics beyond the standard 3-neutrino oscillation framework. The presence of LED modifies the neutrino oscillation. The departure from the standard oscillations can be parameterized by the Dirac mass of the lightest neutrino ($m_0$) and the radius of the largest compactified dimension ($R_{\rm{ED}}$). Various neutrino experiments explore the LED model such as MINOS~\cite{Machado:2011jt,MINOS:2016vvv,Forero:2022skg},  IceCube~\cite{Esmaili:2014esa}, beta decay experiments~\cite{Basto-Gonzalez:2012nel,Rodejohann:2014eka}, reactor neutrino experiments~\cite{DiIura:2014csa,Girardi:2014gna}, short-baseline experiments~\cite{Carena:2017qhd,Stenico:2018jpl}, JUNO~\cite{Basto-Gonzalez:2021aus}, COHERENT measurements~\cite{Khan:2022bcl} etc. In this paper, we predicted the bound on LED parameter space using the proposed experiments T2HK, ESSnuSB, and DUNE. 

The paper is organized as follows: In section~\ref{Formalism}, we show the effect of LED in the lagrangian level, then we diagonalize the mass matrix and provide the expression of the neutrino oscillation probabilities in vacuum and the expression of the neutrino evolution in the presence of matter potential. Section~\ref{simulation_details} provides the simulation details of various experiments. Section~\ref{results} contains the probability, event, and sensitivity analysis. Finally, we conclude in section~\ref{conclusion}. 

\section{Formalism}  
\label{Formalism}
In the LED model, SM particles are confined to four dimensional space, while the SM singlet right handed neutrinos propagate to all dimensions, including the extra dimension. We augment the SM sector with the three  5-dimensional singlet fermions $\Psi^\alpha_{L,R}$ corresponding to three active neutrinos $\nu^{\alpha}_L$. From the four dimensional point of view, these singlet fields can be expressed as a tower of Kaluza-Klein (KK) modes ($\Psi^{\alpha (n)}_{L,R}\, , n=-\infty ..  \infty$) after the compactification of the fifth dimension with a periodic boundary condition. The KK modes behave like a large number of sterile neutrinos. We redefine the fields that couple to SM neutrino as $\nu^{\alpha (0)}_{R} \equiv \Psi^{\alpha (0)}_{R}$ and $\nu^{\alpha(n)}_{L,R} \equiv (\Psi^{\alpha (n)}_{L,R} + \Psi^{\alpha (-n)}_{L,R} )/ \sqrt{2}$. In this basis, the mass term of the Lagrangian~\cite{Davoudiasl:2002fq} after the electroweak symmetry breaking is given by
\begin{eqnarray} 
L_{\rm{mass}} = m^{D}_{\alpha \beta} \big(\bar{\nu}^{\alpha (0)}_{R} \nu^{\beta}_{L} + \sqrt{2} \sum^{\infty}_{n=1} \bar{\nu}^{\alpha (n)}_{R} \nu^{\beta}_{L} \big)
 +  \sum^{\infty}_{n=1} \dfrac{n}{R_{\rm{ED}}} \bar{\nu}^{(n)}_{R} \nu^{(n)}_{L} + h.c. \, ,
 \label{mass_matrix}
\end{eqnarray}
where $m^D$ is the Dirac mass matrix and $R_{\rm{ED}}$ represents the radius of the compactification. The diagonalization of the mass matrix can be achieved in two steps. We define two $3\times 3$ matrices $U$ and $r$ that diagonalize $m^D$ \textit{i.e.} $m^D_{\rm{diag}} = r^{\dagger} m^D U=\rm{diag} (m^D_1, m^D_2, m^D_3)$ and  
\begin{eqnarray}
\nu^{\alpha}_L &=& U^{\alpha i} \nu^{\prime i(0)}_L \\
\nu^{\alpha (n)}_R &=& r^{\alpha i } \nu^{\prime i(n)}_R, n=0...\infty \\
\nu^{\alpha (n)}_L &=& r^{\alpha i} \nu^{\prime i(n)}_L, n=1...\infty .
\end{eqnarray}
In the pseudo mass eigenstates  $\nu^{\prime i}_{L}=\big (\nu^{\prime i},\nu^{\prime i(1)},\nu^{\prime i(2)}, .. \big )^T_{L}  $ and $\nu^{\prime i}_{R}=\big (\nu^{\prime i(0)},\nu^{\prime i(1)},\nu^{\prime i(2)}, .. \big )^T_{R}  $, the mass term in Eq.~\ref{mass_matrix} becomes 
\begin{eqnarray}
L_{\rm{mass}} =\sum^3_{i=1} \bar{\nu}^{\prime i}_R M^i \nu^{\prime i}_L+ h.c.
 \label{mass_matrix1}
\end{eqnarray}
where $M_i$ is an infinite dimensional matrix expressed as
\begin{eqnarray}
M^i= \dfrac{1}{R_{\rm{ED}}}
\begin{pmatrix}
m_i^DR_{\rm{ED}}&0&0&0&\ldots\\
\sqrt{2}m_i^D R_{\rm{ED}}&1&0&0&\ldots\\
\sqrt{2}m_i^DR_{\rm{ED}}&0&2&0&\ldots\\
\vdots&\vdots&\vdots&\vdots&\ddots
\end{pmatrix}
\end{eqnarray}
The true mass basis can achieved via diagonalization of the infinite-dimensional matrix $M^i$. We consider two infinite-dimensional matrices ($L$ and $R$) such that $R^{\dagger}_i M^i L_i$ is diagonal and the actual mass basis is given by $\nu^i_L = L^{\dagger} \nu^{\prime i}_L $ and $\nu^i_R = R^{\dagger} \nu^{\prime i}_R $. Hence, we can write the flavor basis brane neutrino as
\begin{eqnarray}
\nu^{\alpha}_L = \sum^{3}_{i=1} U^{\alpha i} \sum^{\infty}_{n=0} L^{0n}_i \nu^{i(n)}_L.
\end{eqnarray}
$L$ can be determined by the diagonalization of the Hermitian matrix $M^{\dagger} M$~\cite{Arkani-Hamed:1998wuz,Dienes:1998sb,Dvali:1999cn,Barbieri:2000mg} and given by
\begin{eqnarray}
\left(L^{0n}_i\right)^2=\frac{2}{1+\pi^2\left(m_i^{D}R_{\rm{ED}}\right)^2+\left(\lambda^{(n)}_i\right)^2/\left(m_i^{D}R_{\rm{ED}}\right)^2}
\end{eqnarray}
where $\left(\lambda^{(n)}_i\right)^2$ represent the eigenvalues of the matrices $R^2_{\rm{ED}} M^{\dagger}_i M_i$, which can be found from the following equation:

\begin{equation}\label{trans}
\lambda_i^{(n)}-\pi \big(m_i^DR_{\rm{ED}}\big)^2\cot\left(\pi\lambda_i^{(n)}\right)=0.
\end{equation}
The mass of $\nu^{i(n)}_L$ is $\lambda^{(n)}_i/R_{\rm{ED}}$ and 
\begin{eqnarray}
L^{jn}_i = \dfrac{\sqrt{2} j m^D_i R_{\rm{ED}}}{ (\lambda^{(n)}_i)^2-j^2} L^{0n}_i,
\end{eqnarray}
where $j=1..\infty$ and $n=0..\infty$. We are interested in the scenario where the effect of LED can be tread as perturbation on the top of the standard oscillation. This implies that $m^D_i R_{\rm{ED}}<<1$. With this assumption, we can write
\begin{eqnarray}\nonumber
\lambda^{(0)}_i = m^D_i R_{\rm{ED}} \big(1-\dfrac{\pi^2}{6}(m^D_i R_{\rm{ED}})^2+..\big), ~~~~~~~ \lambda^{(j)}_i = j+\dfrac{1}{j}(m^D_i R_{\rm{ED}})^2+..\\ \nonumber
L^{00}_i = 1-\dfrac{\pi^2}{6}(m^D_i R_{\rm{ED}})^2+..,~~~~~~~~~~~~~~~~~~~ L^{0j}_i= \dfrac{\sqrt{2} m^D_i R_{\rm{ED}}}{j} + .. ~~~~~~ \\
L^{j0}_i= -\dfrac{\sqrt{2} m^D_i R_{\rm{ED}}}{j} + .., ~~~~~~~~~~~~~~~~~~~~~~~~~~ L^{jj}_i=1- \dfrac{( m^D_i R_{\rm{ED}})^2}{j^2} + ..,~
\label{lambda_L}
\end{eqnarray}
and $L^{kj}=\mathcal{O}(( m^D_i R_{\rm{ED}})^2)$ for $k \neq j=1..\infty$.
The transition probability of a particular neutrino flavor $\nu_{\alpha}$ to $\nu_{\beta}$ in the presence of LED is given by
\begin{eqnarray}
P_{\alpha\beta}(L,E) = \big |\sum^{3}_{i=1} U^{\alpha i} U^{*\beta i } A_i(L,E)\big |^2,
\label{CC_p}
\end{eqnarray}
where $L$ is the distance between source and detector, $E$ represents the energy of the neutrino, and
\begin{eqnarray}
A_i(L,E) = \sum^{\infty}_{n=0} \left(L^{0n}_i\right)^2 {\rm{exp}} \left(i \dfrac{\lambda_i^{(n)2} L}{2ER^2_{\rm{ED}}} \right).
\label{CC_p_e}
\end{eqnarray}
Similarly, the active to sterile KK modes oscillation probability is 
\begin{eqnarray}
P_{\alpha s} (L,E) =  \sum^{3}_{i=1} \sum^{\infty}_{j=1} \big |B_{\alpha i (j)}\big |^2, 
\label{NC_p} 
\end{eqnarray}
where
\begin{eqnarray}
B_{\alpha i (j)} = U^{\alpha i} \sum^{\infty}_{n=0}  L^{0n}_i L^{jn} {\rm{exp}} \left(i \dfrac{\lambda_i^{(n)2} L}{2ER^2_{\rm{ED}}} \right).
\label{NC_p_e}
\end{eqnarray}
The neutrino masses ($\lambda^{(0)}_i/R_{\rm{ED}}$) of the mostly active neutrinos  and Dirac masses ($m^D_i$) are related by first term of Eq.~\ref{lambda_L} and we can write $\Delta m^2_{ij} R^2_{\rm{ED}} = (\lambda^{(0)}_i)^2 - (\lambda^{(0)}_j)^2 $. The known values of the solar ($\Delta m^2_{21}$) and atmospheric ($\Delta m^2_{31}$) mass squared difference can be utilized to remove two parameters ($m^D_2, m^D_3$) from the theory and the oscillation probability depends only on $m^D_1~(\equiv m_0)$ and $R_{\rm{ED}}$ parameters. 

The vacuum neutrino oscillation probability is modified in matter. In the presence of LED, the time evolution of the neutrino is described by the following equation~\cite{Berryman:2016szd}:
\begin{eqnarray}\label{evolution}
i\frac{d}{dt}{\nu^{\prime}_i}_L=\Bigg[\frac{1}{2E_\nu}M_i^{\dagger}M_i{\nu^{\prime}_i}_L+\sum_{j=1}^3
\begin{pmatrix}
V_{ij} & 0_{1\times n} \\
0_{n\times 1} & 0_{n\times n}
\end{pmatrix}
{\nu^{\prime}_i}_L\Bigg]_{n\to\infty},~ V_{ij}=\sum_{\alpha=e,\mu,\tau} U^*_{\alpha i}U_{\alpha j}\Big(\delta_{\alpha e}V_{\rm{CC}}+V_{\rm{NC}}\Big),\nonumber\\
\end{eqnarray}
where the $V_{CC}=\sqrt{2}G_Fn_e$ and $V_{NC}=-1/\sqrt{2}\, G_Fn_n$ are the charged and neutral current matter potential, respectively. $n_e$ and $n_n$ represent the number density of electron and neutron, respectively. We consider the equal number density of electron and neutron and keep the matter density constant throughout the neutrino evolution for different baselines. We assume two KK modes for our analysis, and the result remains unaffected by the inclusion of the higher number of modes.

\section{Simulation Details}
\label{simulation_details}
\textit{T2HK:} There is a plan to upgrade the Super-Kamiokande (SK) \cite{Super-Kamiokande:1998kpq} program in Japan to Hyper-Kamiokande (HK) \cite{Hyper-KamiokandeProto-:2015xww,Hyper-Kamiokande:2016srs,Hyper-Kamiokande:2018ofw}. This involves a roughly 20-fold increase in the fiducial mass of SK. We consider two 187 kt third-generation Water Cherenkov detectors that will be installed close to the existing SK site. The detector will also receive an intense beam of neutrinos form the J-PARC proton accelerator research complex in Tokai, Japan, which is placed 295 km away from the detector site. This setup is commonly known as T2HK. The proton beam power at the J-PARC facility is 1.3 MW which will generate $27\times10^{21}$ protons on target (P.O.T.) in  a total run time of 10 years.
We consider the 2.5$^{\circ}$ off-axis flux and the run time is divided in a 1:3 ratio in neutrino and anti-neutrino modes, \textit{i.e.}, 2.5 years for the neutrino run while 7.5 years for the anti-neutrino run. The signal normalization errors in the $\nue(\anue)$ appearance and $\numu(\anumu)$ disappearance channels are 3.2\% (3.6\%) and 3.9\% (3.6\%), respectively. In this work, we consider the background and energy calibration errors to be 10\% and 5\%, respectively, for all channels.
\vspace{0.2cm}

\textit{ESSnuSB:} ESSnuSB~\cite{ESSnuSB:2013dql,Alekou:2022emd,Ghosh:2021fua} is a proposed water Cherenkov detector with a fiducial mass of 538 kt. There are two possible baseline configurations for the setup: either at 540 km or at 360 km away from the source. European Spallation Source (ESS) will produce an intense neutrino flux by hitting $2.7 \times 10^{23}$ protons on target per year. The kinetic energy of the proton is 2.5 GeV.
We consider an overall $5\%$ signal
and $10\%$ background normalization errors both in the appearance and disappearance channels for the neutrino and anti-neutrino runs. The ten years of run-time are equally divided between the neutrino and anti-neutrino modes. We consider the same setup configuration for both baselines in our analysis. 
\vspace{0.2cm}

\textit{DUNE:} DUNE~\cite{DUNE:2015lol,DUNE:2016rla} is an
upcoming superbeam long-baseline (1300 km) neutrino experiment at Fermilab, U.S.A., capable of determining the present unknowns of the neutrino oscillation parameters. The optimized beam of 1.07 MW - 80 GeV protons at Fermilab will provide $1.47\times 10^{21}$ P.O.T. per year. We consider a 40kt Liquid Argon (LAr) detector and a total of 7 years of run-time, which is equally divided between neutrinos and anti-neutrinos runs. All the experimental details are taken from \cite{DUNE:2016ymp}. We consider both the charged and neutral current measurements in our analysis. The detailed information for the NC events is taken from \cite{LBNE:2013dhi}. The detection efficiency of the NC event is assumed to be 90$\%$. In a NC process, the outgoing neutrino also takes away some fraction of the incoming neutrino energy. Because of this, the reconstructed energy is typically lower than the total incoming energy and a gaussian energy resolution function cannot provide the correct events spectra. We use migration matrices \cite{DeRomeri:2016qwo} to simulate the NC events spectra appropriately. Moreover, we also take 10$\%$ of the CC events as NC background, both in the neutrino and anti-neutrino modes. The normalization errors for the signal and background are assumed to be 5$\%$
and 10$\%$, respectively. All other pertinent details regarding the NC analysis are taken from \cite{Gandhi:2017vzo}.
\vspace{0.2cm}

We use GLoBES software~\cite{Huber:2004ka,Huber:2007ji} to simulate the events in various detectors. The effect of LED is included by changing the probability engine. We consider the constant matter density ($2.95$ gm/cc) for our simulation purposes. The value of standard oscillation parameters and its marginalization range are  given in Tab. \ref{Best-fit}. We use the central values of the oscillation parameters for the probability, event, and sensitivity predictions in the next section.  

\begin{table}[ht]
\centering
\begin{tabular}{|| c || c | c | }
\hline
Parameter & Normal Hierarchy (NH) & Inverted Hierarchy (IH) \\
\hline \hline
$\sin^2\theta_{12}$ & $0.304\pm 0.012$ & $0.304\pm 0.012$ \\
\hline
$\sin^2\theta_{13}$ & $0.02219 \pm 0.00062$ & $0.02219 \pm 0.00062$ \\
\hline
$\sin^2\theta_{23}$ & $0.5 \pm 0.086$ & $0.5 \pm 0.086$ \\
\hline
$\Delta m_{21}^2$ & $(7.42\pm 0.02)\times 10^{-5}$ eV$^2$ & $(7.42\pm 0.02)\times 10^{-5}$ eV$^2$ \\
\hline
$\Delta m_{31}^2$ & $(2.517\pm 0.028 )\times 10^{-3}$ eV$^2$ & $-(2.498\pm 0.028)\times 10^{-3}$ eV$^2$ \\
\hline
$\delta_{cp}$ & $-\pi/2 \,\, [-\pi : \pi]$  & $-\pi/2\,\, [-\pi : \pi]$ \\
\hline
\end{tabular}
\caption{\em Benchmark value and its marginalization range for the standard neutrino oscillation parameters for the normal and inverted hierarchy scenarios. These values are compatible with the global fit of the oscillation parameters~\cite{Esteban:2020cvm,Capozzi:2021fjo,deSalas:2020pgw}.}
\label{Best-fit}
\end{table}

\begin{figure*}[]
    \centering
    \includegraphics[width = 0.49 \textwidth]{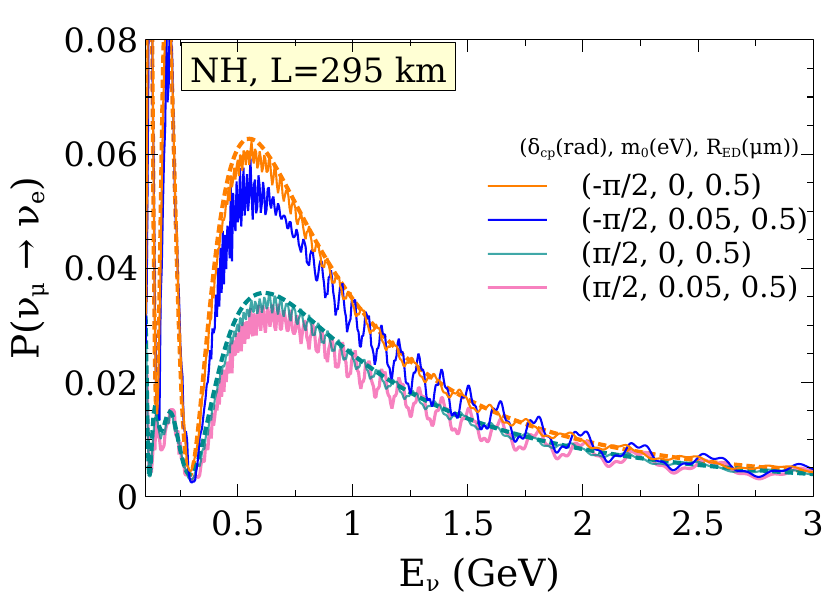}
    \includegraphics[width = 0.49\textwidth]{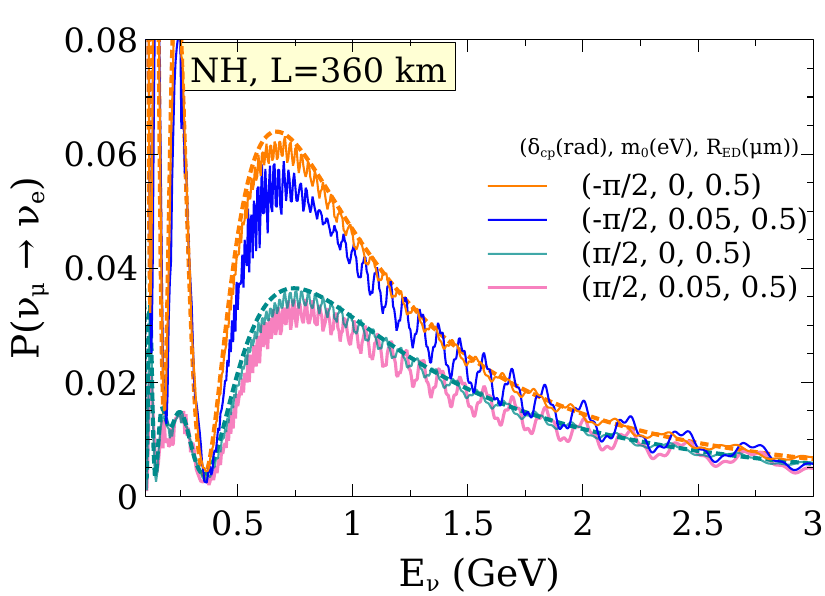}
     \includegraphics[width = 0.49 \textwidth]{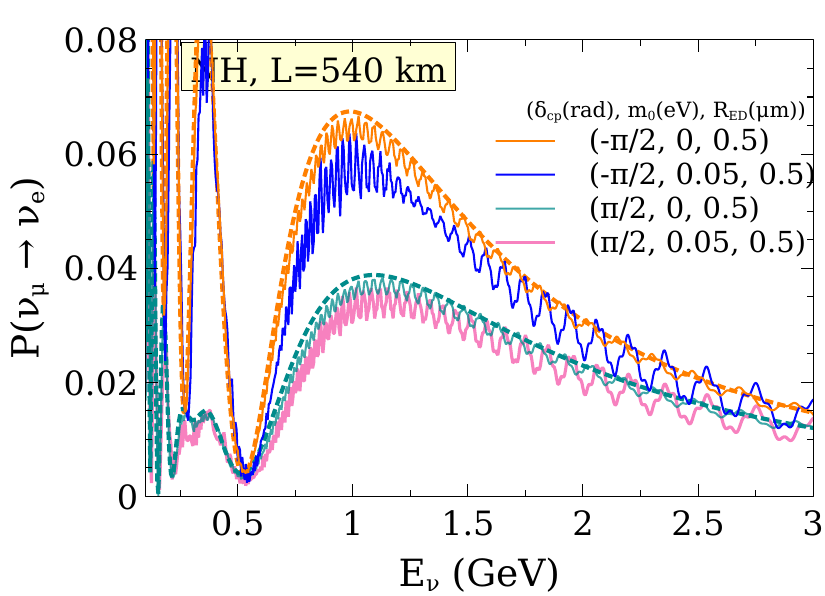}
    \includegraphics[width = 0.49\textwidth]{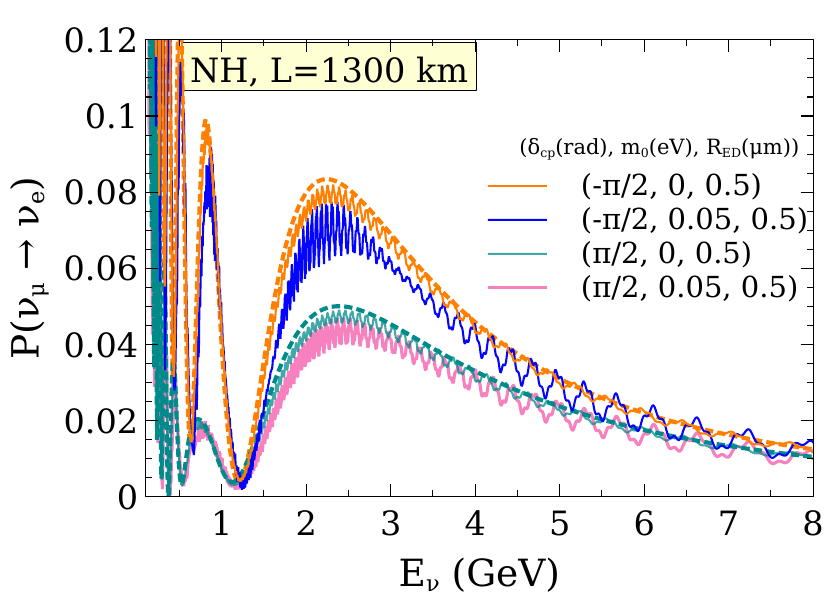}
   \caption{\em Appearance oscillation probabilities in the presence of the LED for two values of CP phases and baselines. The dotted line corresponds to the standard oscillation scenario.}
   \label{Fig.Probability}
\end{figure*}

\section{Results}
\label{results}
\begin{figure*}[]
    \centering
    \includegraphics[width = 0.49 \textwidth]{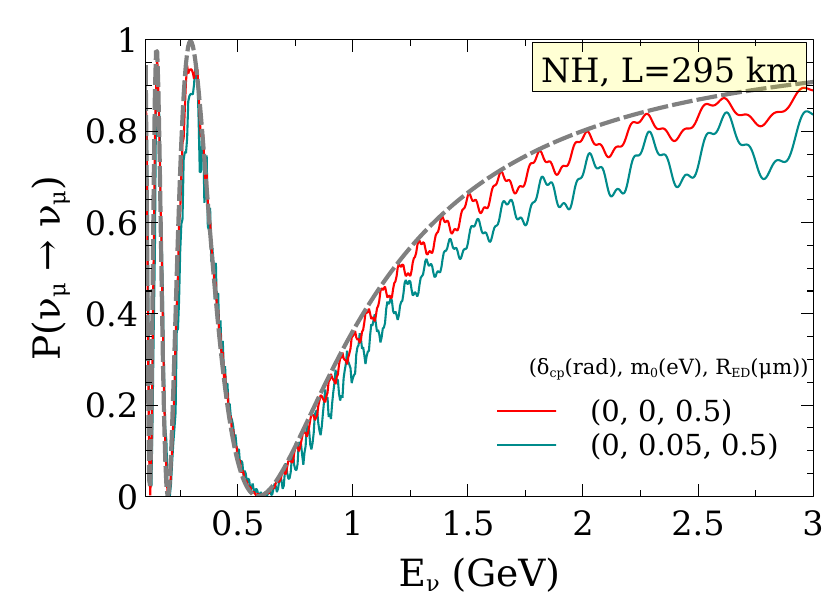}
    \includegraphics[width = 0.49\textwidth]{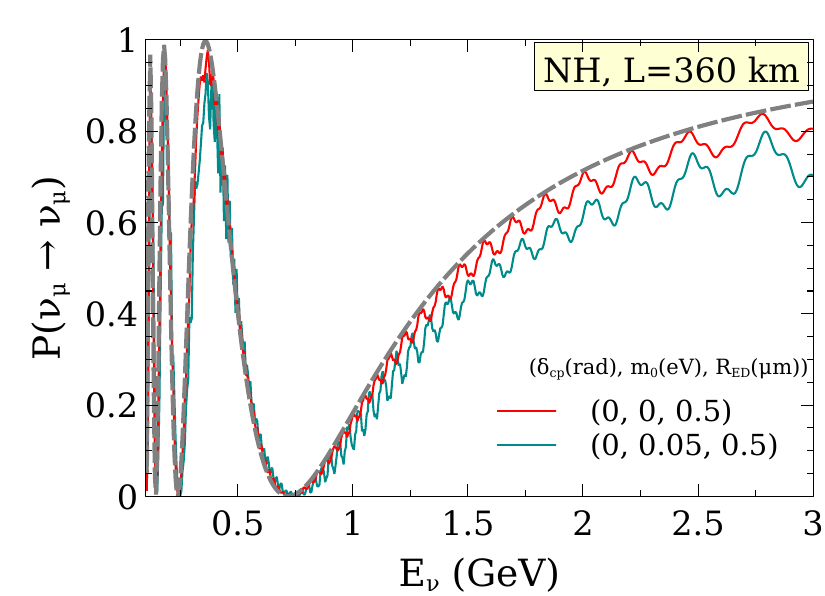}
     \includegraphics[width = 0.49 \textwidth]{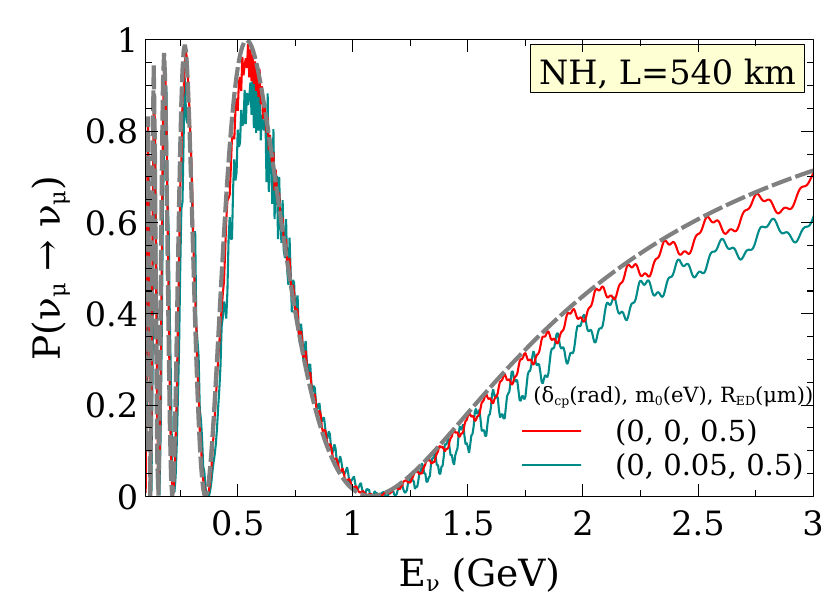}
    \includegraphics[width = 0.49\textwidth]{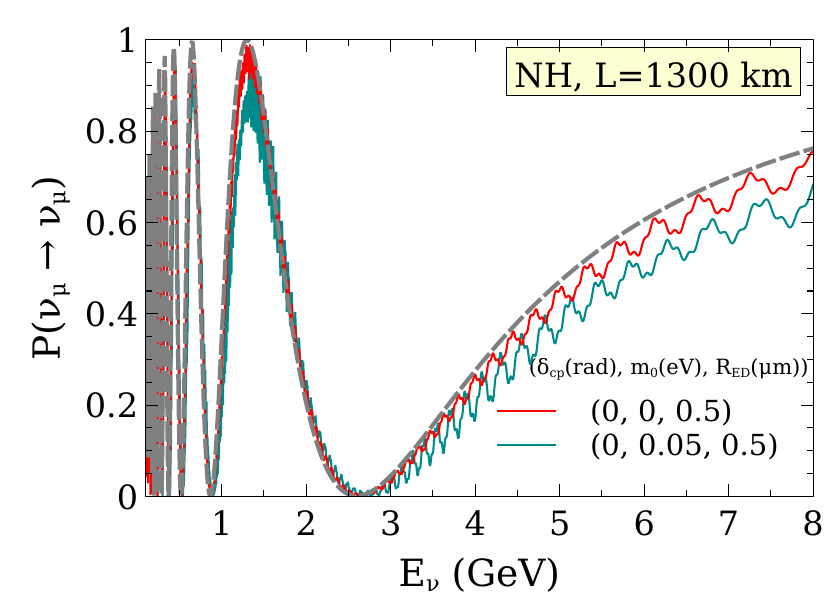}
   \caption{\em Disappearance oscillation probabilities in the presence of the LED for different baselines and LED parameters. We choose the value of $\delta_{cp}$ to be 0, and all other CP phases will give almost similar probabilities as the dependence of the CP phase on the disappearance channel is mild. The dotted line corresponds to the standard scenario.}
   \label{Fig.ProbabilityPmm}
\end{figure*}

\begin{figure*}[tbh!]
    \centering
    \includegraphics[width = 0.49 \textwidth]{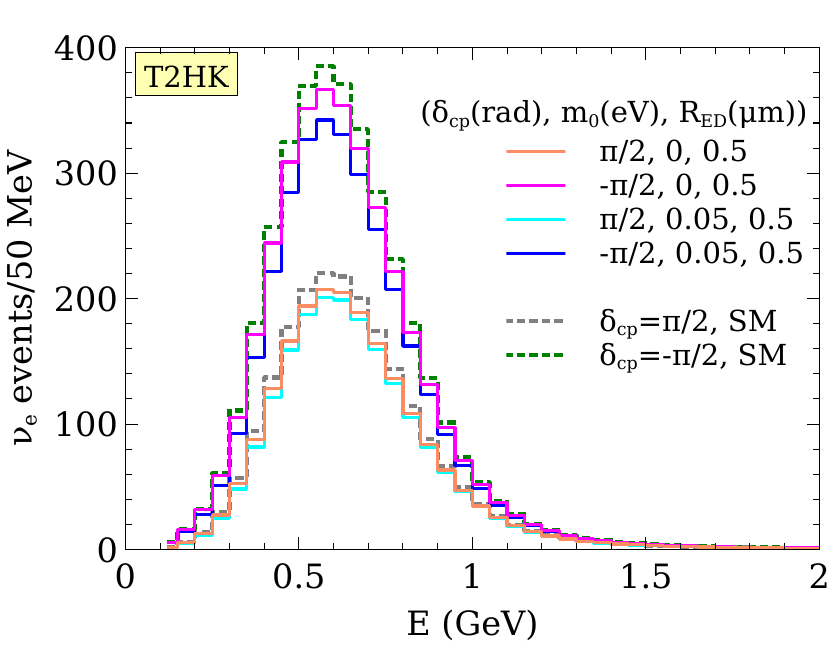}
    \includegraphics[width = 0.49\textwidth]{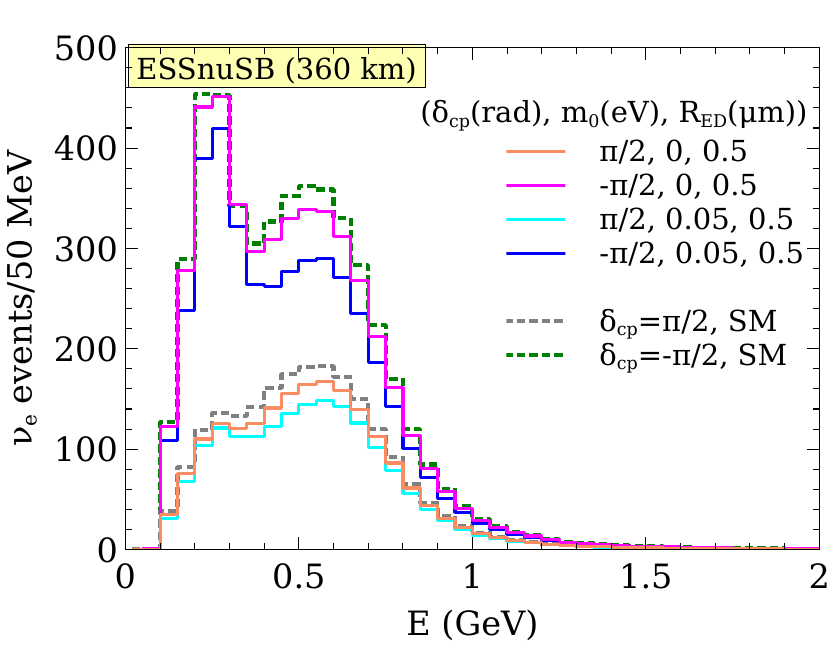}
     \includegraphics[width = 0.49 \textwidth]{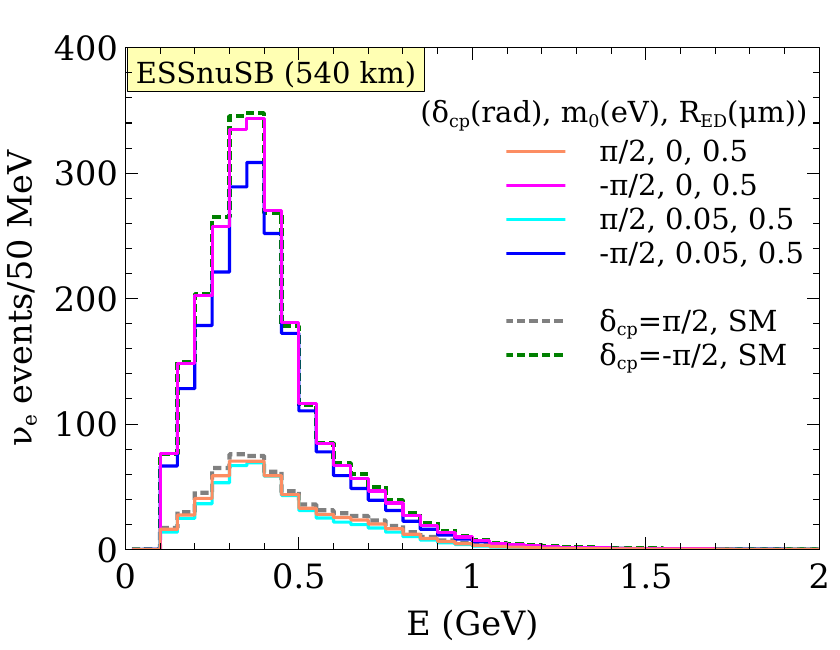}
    \includegraphics[width = 0.49\textwidth]{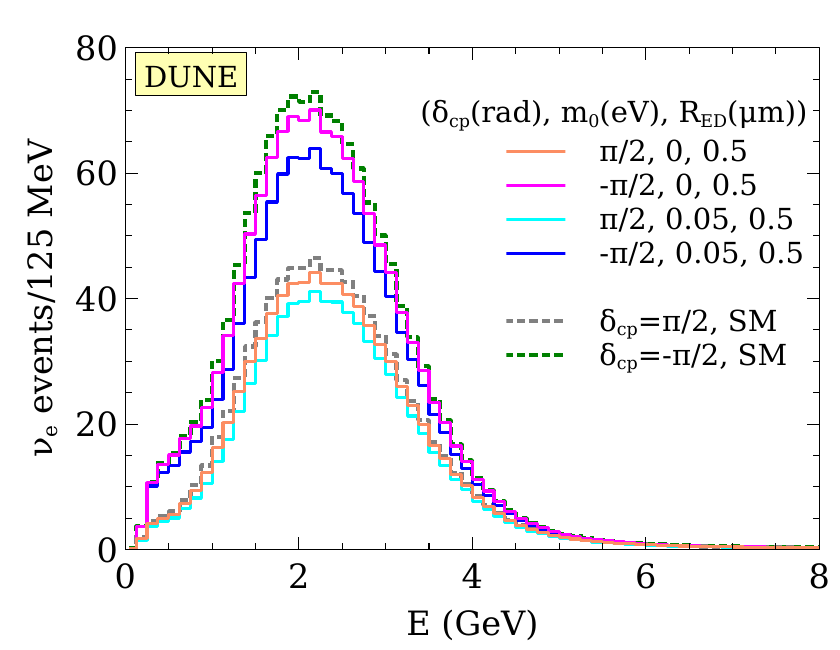}
   \caption{\em Appearance events in T2HK, ESSnuSB, and DUNE detectors for the standard (dotted lines) as well as the LED case (solid lines) for various values of CP phase. }
   \label{Fig.Eventsnue}
\end{figure*}

\begin{figure*}[tbh!]
    \centering
    \includegraphics[width = 0.49 \textwidth]{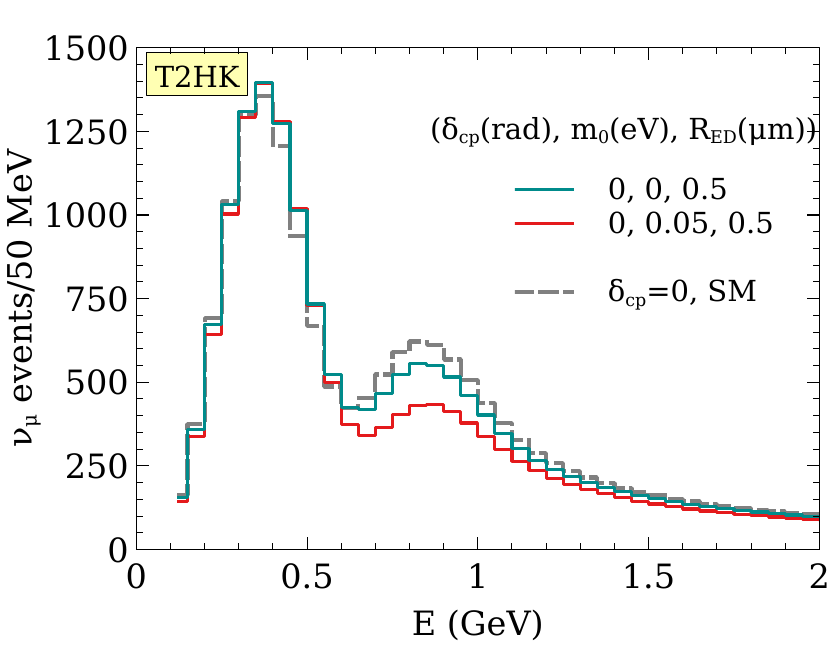}
    \includegraphics[width = 0.49\textwidth]{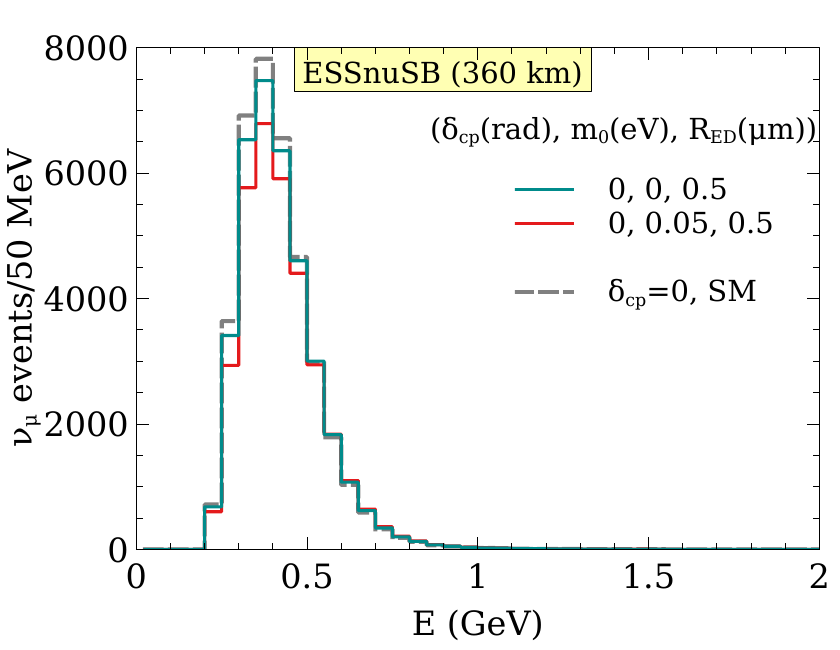}
     \includegraphics[width = 0.49 \textwidth]{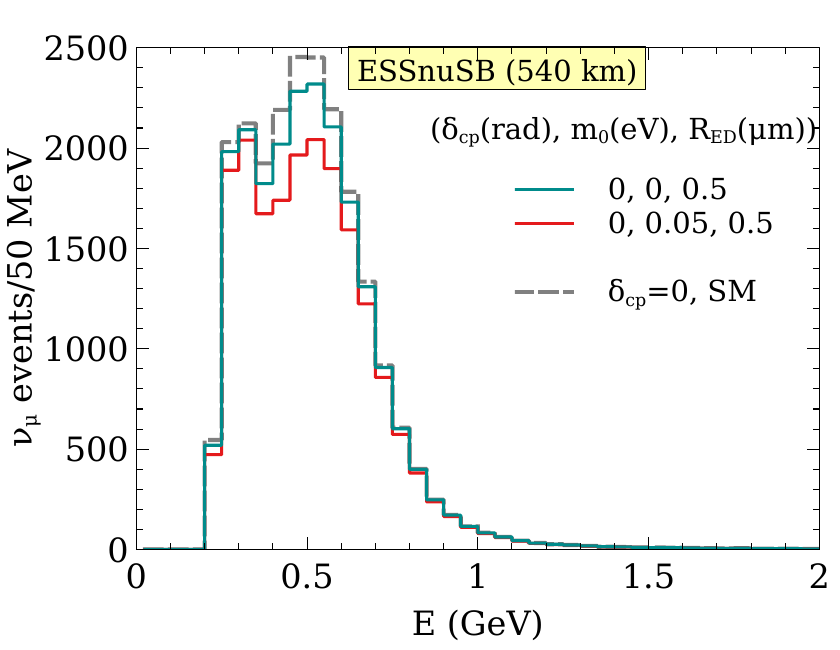}
    \includegraphics[width = 0.49\textwidth]{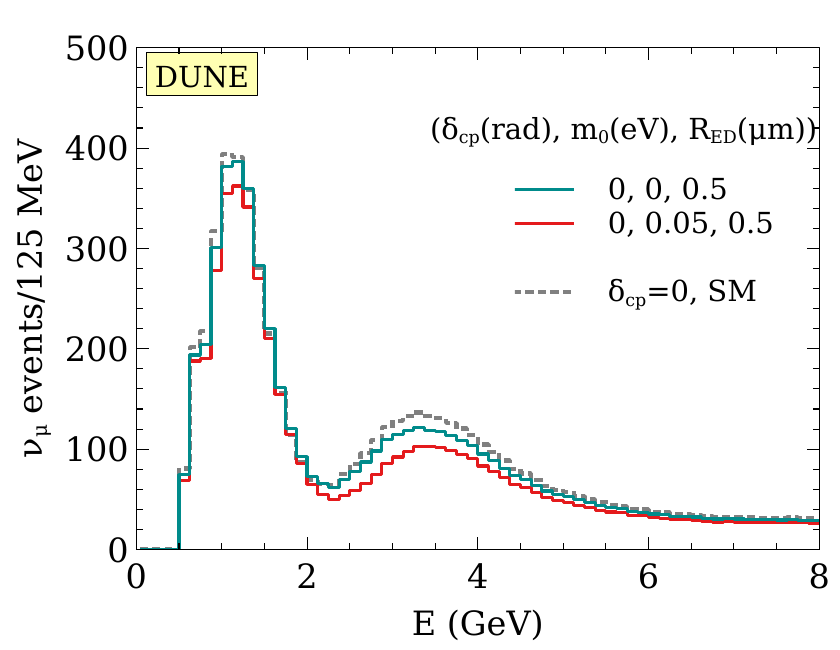}
   \caption{\em Disappearance events in T2HK, ESSnuSB, and DUNE detectors for the standard (dotted lines) as well as the LED case (solid lines) for $\delta_{cp}=0$.}
   \label{Fig.Eventsnumu}
\end{figure*}

In Fig.~\ref{Fig.Probability}, we show the neutrino appearance oscillation probabilities for different baselines, \textit{i.e.}, T2HK (295 km), ESSnuSB (360 km), ESSnuSB (540 km), and DUNE (1300 km). The dotted line corresponds to the standard neutrino oscillation for the two choices of $\delta_{cp}$ $(-\pi/2$ and $\pi/2)$, while the solid line represents the probability with LED model for two combinations of $(m_0, \, R_{\rm{ED}})$ parameters for a given value of CP phase. We can observe from the plot that the oscillation probability increases with the increase in baseline length due to the increasing matter effect and the first oscillation maxima shifting towards higher neutrino energies. In the presence of LED, the oscillation probabilities decrease from the standard prediction, and wiggle appears due to the fast oscillation in the KK modes. The oscillation pattern remains almost similar for all the baselines. The effect of the LED increases with the increase of the lightest Dirac mass for a particular value of $R_{\rm{ED}}$. This is also evident from Eqs.~(\ref{lambda_L}, \ref{CC_p}). In Fig.~\ref{Fig.ProbabilityPmm}, we show the disappearance probability for different baselines for the standard (dotted lines) as well as LED cases (solid lines). We show the plot only for one value of the CP phase ($\delta_{cp}=0$), as the effect of the CP phase is mild in the disappearance channel and all other values of the CP phase give almost similar probabilities. The deviation from the standard oscillation due to LED becomes larger with higher incoming neutrino energy beyond the first oscillation maxima point. Hence, if the neutrino fluxes peak at a higher energy than the first oscillation maxima, then the effect of LED will be greater. The appearance and disappearance events for the neutrino run are shown in Fig.~\ref{Fig.Eventsnue} and Fig.~\ref{Fig.Eventsnumu}, respectively. For T2HK and DUNE, the maximum number of appearance events occurs where both the first oscillation maxima and the peak of the fluxes coincide. But that is not the case for the ESSnuSB detector at 540 km. The second oscillation maxima and peak of the fluxes appear at $\sim 0.3$ GeV, and there are almost negligible fluxes beyond $1$ GeV. Also, for the ESSnuSB detector at 360 km, the peak of the flux and the first oscillation maxima do not coincide. We can observe from the plot that the standard and LED cases $\nu_e$ events change significantly with the change of $\delta_{cp}$ values. In the disappearance channel, we only show the $\delta_{cp}=0$ case, as the dependence of events on the CP phase is very small. The deviation of the events from the standard scenario is greater for T2HK and DUNE, where there are sufficient fluxes beyond the first oscillation maxima. For the ESSnuSB detector, most of the events are concentrated around 0.5 GeV, and there are almost negligible events beyond 1 GeV. The number of events decreases from the standard prediction in the presence of the LED, both in the appearance and disappearance channels. To quantify the effects of the LED parameters on standard oscillation, we perform the $\chi^2$ analysis next.

The Poissonian $\chi^2$ is defined as \cite{Huber:2002mx}
\begin{equation}
\chi^2 = \min_{\vec{\lambda},\alpha,\beta}\big[~\sum^n_{i=1} 2\big( N^{test}_i -N^{true}_i + N^{true}_i \,\,ln ~\dfrac{N^{true}_i}{N^{test}_i} {\big)} + \big(\dfrac{\alpha}{\sigma_s}\big)^2 + \big(\dfrac{\beta}{\sigma_b}\big)^2 ~\big],
\end{equation}
where $\vec{\lambda}=\{\Delta m^2_{21},  \Delta m^2_{31}, \theta_{12}, \theta_{13}, \theta_{23},  \delta_{cp}\}$, and $\alpha$, $\beta$ are two nuisance parameters. $\sigma_s$ and $\sigma_b$ represent the signal and background normalization errors, respectively. $N^{test}_i$ and $N^{true}_i$ are the test and true  data sets in the \textit{i-th} energy bin. $N^{true}_i$ can be expressed as
\begin{eqnarray}
N^{true}_i = s_{i}(\vec{\lambda})(1+\alpha) + b_{i}(\vec{\lambda})(1+\beta),
\end{eqnarray}
where $s_i$ and $b_{i}$ represent the signal and background events in the \textit{i-th} energy bin. $N^{test}_i = s_{i}(\vec{\lambda},m_0, {\rm{R_{\rm{ED}}}})+ b_{i}(\vec{\lambda},m_0, {\rm{R_{\rm{ED}}}})$. We generate the true data set using the standard oscillation scenario with the central values of the oscillation parameters as in Tab.~\ref{Best-fit}. In the test data sets, we consider the LED model and marginalize the $\chi^2$ over both the oscillation parameter uncertainties and  systematic uncertainties, and report the minimum $\chi^2$.
 
\begin{figure*}[tbh!]
    \centering
    \includegraphics[width = 0.49 \textwidth]{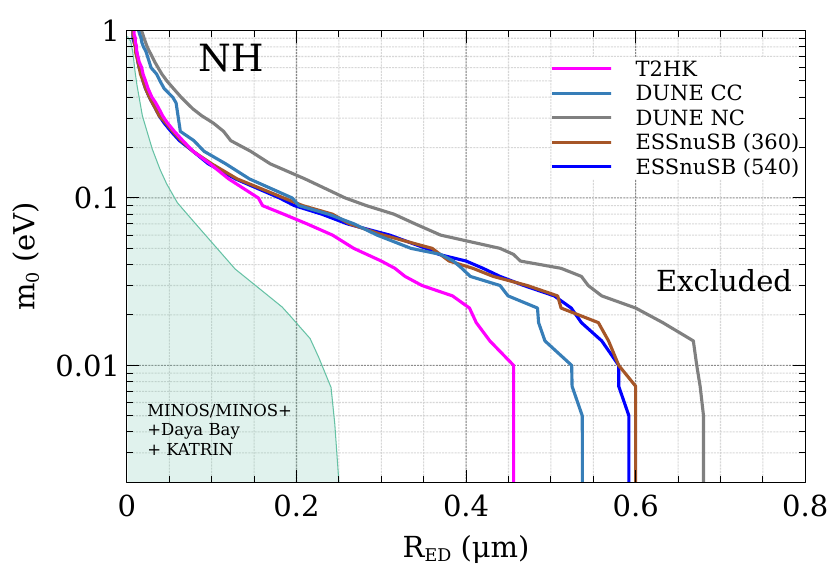}
    \includegraphics[width = 0.49\textwidth]{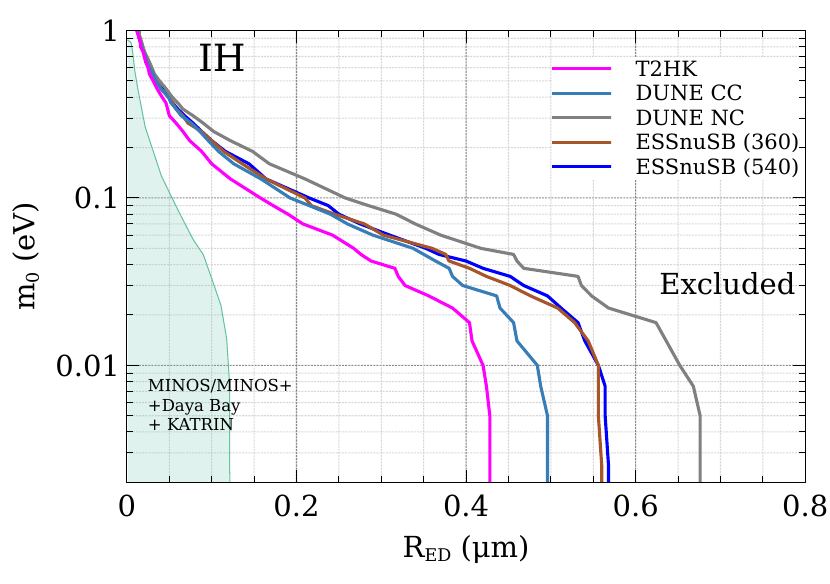}
   \caption{\em The constraints on LED parameters in the $R_{\rm{ED}}-m_0$ plane at $90\%$ C.L. ($\Delta \chi^2 = 4.61$) for the normal (left panel) and inverted (right panel) hierarchy scenarios. The green shaded region represents the constraint coming from the combined analysis of MINOS/MINOS+, Daya Bay, and KATRIN~\cite{Forero:2022skg}.}
   \label{Fig.m0_Red}
\end{figure*}

The bounds on LED parameters at $90\%$ C.L. ($\Delta \chi^2 = 4.61$, two degrees of freedom) in the $R_{\rm{ED}}-m_0$ plane are shown in Fig.~\ref{Fig.m0_Red}. The left and right panels correspond to the normal and inverted hierarchy scenarios, respectively. Regions towards the right (left) of the curve are excluded (allowed) at $90\%$ C.L. from the respective experiments. We can observe from the figure that T2HK will be able to provide the best bound on the LED parameters, and at $90\%$ C.L., it can exclude the value of $R_{\rm{ED}}>0.45 ~\mu $m ($R_{\rm{ED}}>0.425 ~\mu $m) for the NH (IH) scenario. The better constraint in T2HK is attributed to the higher statistics compared to all other experiments. This is also seen from the event plots in Figs.~(\ref{Fig.Eventsnue}, \ref{Fig.Eventsnumu}). With the increase in mass $m_0$, the constraint on $R_{\rm{ED}}$ becomes more stringent. At $m_0=1$ eV, all the experiments can rule out $R_{\rm{ED}}\gtrsim 0.02 ~\mu $m. The constraint on the extra dimension using the charged current measurements is shown by the cyan line for DUNE. These results are consistent with \cite{DUNE:2020fgq,Berryman:2016szd}. Here, we also explore the capabilities of neutral current measurements. The NC events depend on the total number of active flavors present in the neutrino beam. Due to the oscillation in KK sterile modes, the total number of active flavors drops from unity depending on the values of $m_0$ and $R_{\rm{ED}}$. We can see from Fig.~\ref{Fig.m0_Red} that the NC measurements provide much weaker constraints compared to the CC measurements. We can understand this from Eqs.~(\ref{NC_p}, \ref{NC_p_e}). Due to the presence of $L^{jn}$ terms, the dependence of NC probabilities on $m_0$ and $R_{\rm{ED}}$ is much weaker than CC oscillation probabilities. We also checked that the combination of CC and NC did not improve the results further. 
The two baseline configurations of ESSnuSB will provide almost similar bounds on the LED parameters for the NH and IH scenarios. ESSnuSB can rule out the $R_{\rm{ED}}>0.60 ~\mu $m ($R_{\rm{ED}}>0.55 ~\mu $m) for the NH (IH) scenario. The present constraint on the LED model coming from MINOS is $R_{\rm{ED}}>0.7~\mu$m~\cite{MINOS:2016vvv} for a very small mass of $m_0$ in the NH scenario. MINOS/MINOS+ and Daya Bay~\cite{Forero:2022skg} rule out $R_{\rm{ED}}>0.25~(0.29)~\mu $m and $R_{\rm{ED}}>0.65~(0.12)~\mu $m, respectively, for NH (IH). For MINOS (MINOS+), the fluxes peaked at 3 GeV (7 GeV), while the first oscillation maxima was at 1.4 GeV. We observe from the disappearance plot that the deviation from the standard prediction increases for larger neutrino energies beyond the first oscillation maxima point. Due to the availability of higher energy fluxes, the constraint coming from MINOS (MINOS+) is more stringent than T2HK, DUNE, and ESSnuSB. Daya Bay experiment is capable of putting a strong bound on LED parameters for IH compared to NH as the $\bar{\nu}_e$ disappearance channel is affected more by LED parameters in the IH scenario. The absolute mass of the neutrino is constrained by the KATRIN experiment. Hence, the combined experiments (MINOS/MINOS+, Daya Bay, and KATRIN) provide strong constraints on the LED parameters~\cite{Forero:2022skg} as shown by the green shaded region in Fig.~\ref{Fig.m0_Red}. Icecube data excludes $R_{\rm{ED}}>0.4~\mu $m~\cite{Esmaili:2014esa} at $2\sigma$ C.L. The future long-baseline experiments T2HK, ESSnuSB, and DUNE will be able to test the LED model independently, and the constraints are comparable to the existing bounds on $R_{\rm{ED}}$ parameters. These bounds are two orders of magnitude stronger than the constraints coming from tabletop experiments, which put a bound on $R_{\rm{ED}}>37~\mu$m~\cite{ParticleDataGroup:2020ssz} at $95\%$ C.L.

\section{Conclusions}
The LED model provides an attractive solution to the hierarchy problem. It also explains the small neutrino mass in a natural way. In this model, all the SM fields are confined to 4-dimensional space, and SM singlet right handed neutrinos could propagate in more than four-dimensional space. The large volume of the extra dimension provides suppression of the coupling of right handed neutrino to 4-dimensional SM neutrino fields and generates small neutrino mass. We consider three 5-dimensional right handed neutrino fields. These fields behave as a tower of KK modes in 4-dimensional space after the compactification of the fifth dimension. The oscillation probability depends on the value of the lightest Dirac mass ($m_0$) and the value of the compactification radius ($R_{\rm{ED}}$). We investigate the capability of the proposed long-baseline experiments T2HK, ESSnuSB, and DUNE to explore the LED parameter space. We find that T2HK will provide the most stringent constraint on $R_{\rm{ED}}$ compared to ESSnuSB and DUNE. We show the capability of  NC measurements to constrain the LED parameters at DUNE. The constraint coming from NC measurements is weaker compared to CC measurements. The combination of CC and NC will not improve the bounds further. The two baseline configurations of ESSnuSB able to give almost similar constraints on $R_{\rm{ED}}$ for the NH and IH scenarios. 
\label{conclusion}

\section*{Acknowledgements} 
We thank Monojit Ghosh for providing the GLoBES glb file for the ESSnuSB detector on behalf of the ESSnuSB collaboration. This work was partially supported by the research grant number 2017W4HA7S ``NAT-NET: Neutrino and Astroparticle Theory Network'' under the program PRIN 2017 funded by the Italian Ministero dell'Universit\`a e della Ricerca (MUR).

\bibliography{Bibliography}
\end{document}